\documentclass[reprint, aps, superscriptaddress]{revtex4-2}

\usepackage{graphicx}
\usepackage{amsmath}
\usepackage{amssymb}
\usepackage{dcolumn}% Align table columns on decimal point
\usepackage{bm}% bold math
\usepackage{layouts}
\usepackage{siunitx}
\usepackage[colorlinks=true, linkcolor=blue, citecolor=blue, urlcolor=blue]{hyperref}
\usepackage[capitalise,noabbrev,nameinlink]{cleveref}
\usepackage[dvipsnames]{xcolor}

\begin{document}

% \preprint{APS/123-QED}

\title{Dynamically enabled transition pathways in multistable systems}

\author{Franco N. Piñan Basualdo}
  \email{franconicolas.pinanbasualdo@kuleuven.be}
\author{Benjamin Gorissen}%
\affiliation{%
 Department of Mechanical Engineering\\
 Katholieke Universiteit Leuven, 3001, Leuven, Belgium
}

\date{\today}

\begin{abstract}

Systems composed of interacting bistable elements are commonly described by transition graphs that determine which state changes are accessible under an external drive. Under quasistatic loading, accessibility is constrained by the equilibrium structure of the system, often resulting in sparse transition networks and unreachable stable states. Here, we show that dynamic loading of dissipatively-coupled hysteron networks enhances accessibility by enabling transition pathways that are forbidden under quasistatic driving while preserving the underlying equilibrium states. In particular, we consider pulse actuation and derive a control map linking pulse amplitude and duration to state transitions. For suitable dissipative couplings, individual hysterons become independently addressable using a single scalar input, increasing transition-graph connectivity and enabling access to otherwise unreachable states. We validate the framework experimentally using pneumatic hysterons and find good agreement with theory. More generally, the framework applies to dissipatively coupled networks of bistable elements across fluidic, mechanical, and electrical domains.

\end{abstract}

% \keywords{Suggested keywords}

\maketitle

Systems composed of interacting nonlinear elements often possess multiple stable equilibrium states, giving rise to rich behaviors such as memory~\cite{keim2011generic,shohat2022memory,paulsen2025mechanical}, and path‑dependent transitions~\cite{coulais2018multi,omidvar2026racetrack}. Examples include snapping beams~\cite{kwakernaak2023counting}, origami structures~\cite{jules2022delicate}, and shells~\cite{faber2020dome}. Across these diverse platforms, multistability emerges from geometric~\cite{van2023nonlinear} or material nonlinearities~\cite{overvelde2015amplifying} and enables programmable mechanical responses and physical computation~\cite{yasuda2021mechanical}. In particular, the ordering of snap events can be engineered to realize prescribed transition sequences and functionalities~\cite{gorissen2019hardware,melancon2022inflatable, chen2026rotary}. 
A convenient abstraction of such behavior is the hysteron model, which exhibits two stable states in the unloaded configurations. Switching between states occurs when the driving field crosses certain thresholds~\cite{lindeman2025generalizing}. 
The interaction of multiple hysterons in a network leads to rich collective behavior and enables mechanical computation~\cite{bense2021complex,liu2024controlled,shohat2025geometric}. 
Under quasistatic assumptions, the stable states of the system depend on the coupling constraints~\cite{pinan2026equilibrium}, and the transitions between them can be traced following an energy-minimizing trajectory~\cite{muhaxheri2025catastrophic}. This quasistatic discrete response of the system is often represented by a transition graph~\cite{terzi2020state,teunisse2025transition,stinissen2026interacting}. This framework can give rise to complex phenomena, such as avalanches~\cite{jin2025dynamic}, where several elements transition in rapid succession, and Garden-of-Eden states~\cite{muhaxheri2024bifurcations}, which are stable states that cannot be reached from the saturated states.
% Generally, the amount of possible transition graphs increases rapidly with the number of hysterons in the systems~\cite{van2021profusion}, making \BG{we need to have a sentence here why this is not good. Not practical?}

However, in reality, dynamic effects are inherent to physical systems and can temporarily drive the system away from equilibrium and lead to responses that differ from the quasistatic case~\cite{lindeman2023competition}. In most analyses, these dynamic excursions are minimized or neglected to simplify the description of the system. In some cases, snap‑through behavior can inevitably cause dynamic effects even when the driving field is varied quasistatically~\cite{jin2025dynamic}.
More recently, it has been shown that dynamic driving can enable selective state transitions. Independent bit addressability can be attained by exploiting the system’s sensitivity to the velocity and acceleration of the driving signal, defining switching boundaries in a two-dimensional control space~\cite{gutierrez2026dynamic}.
In this work, we show that system dynamics enable history-dependent responses under a single driving signal. Instead of expanding the input dimensionality, independent bit addressability emerges from the temporal evolution of the input mediated by the system dynamics.

\begin{figure}
    \centering
    \includegraphics[scale=1.0]{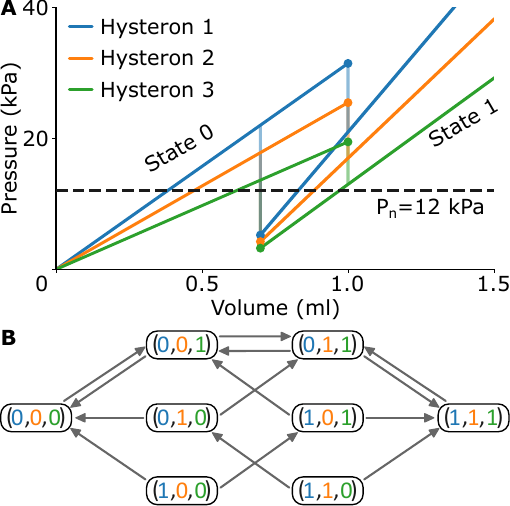}
    \caption{Quasistatic loading of coupled inflatable hysterons. \textbf{(A)} Inflatable hysterons characteristic response. In this case, there exists a neutral pressure $P_n$ at which all collective states are stable. \textbf{(B)} Transition graph under quasistatic loading. In this case, the snap order is given by the order of the pressure peaks and valleys.}
    \label{fig:static}
\end{figure}

As elements, we consider fluidic hysterons, each characterized by a hysteretic pressure–volume response. Each hysteron can then be fully defined by its state $s_i\in\{0,1\}$ and its volume $v_i$ and pressure $p_i$. Unlike standard hysteron models, which consider hysteresis with respect to a single driving field (pressure), we consider fluidic hysterons that show hysteresis both under pressure and volume loading, as shown in \cref{fig:static}A. Snapping under constant volume is limited by internal fluid redistribution and membrane inertia, which occur on timescales much shorter than those associated with network-mediated transport. 
When $N$ such fluidic hysterons are serially connected and loaded with a common pressure input $P_\mathrm{in}$, there is no interaction between the hysterons. We define the system state as $S=\{s_i\}$, its pressure as $P=p_i\,\forall i$, and its volume as $V=\sum_i v_i $. In this case, the snapping order is solely determined by the order of their respective pressure peaks and valleys, which can be condensed in a transition graph as shown in \cref{fig:static}B. If the lowest pressure peak is higher than the highest pressure valley, then there exists a neutral pressure ($P_n$) at which all states $S$ are stable, as shown in \cref{fig:static}A. This transition graph is generally sparsely connected since any state can experience at most two transitions, one when increasing the pressure and one when decreasing it. This often leads to Garden-of-Eden states, which are stable states that are not reachable from the saturated states, e.g., state $(0,1,0)$ in \cref{fig:static}B. 

\begin{figure}
    \centering
    \includegraphics[scale=1.0]{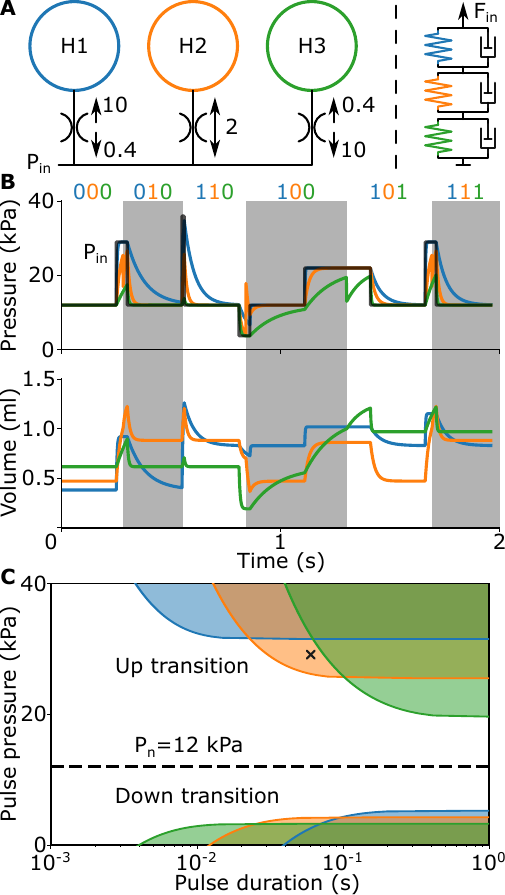}
    \caption{Dynamic loading of coupled fluidic hysterons. \textbf{(A)} Dissipative network with direction-dependent flow restrictors loaded with a dynamic pressure input and the equivalent linear spring system. The conductivity values are in \unit{ml/(s.kPa)}. \textbf{(B)} System response under input pressure pulses. \textbf{(C)} Control map showing the up- and down-transition zones for each hysteron under a pressure pulse as a function of pulse pressure and duration. The cross denotes the parameters of the first pulse in \textbf{(B)}, which induces the up-transition of the second hysteron. The map is computed for the neutral equilibrium condition, $p_i=\SI{12}{kPa}\;\forall i$.}
    \label{fig:dynamic}
\end{figure}

To endow the system with controllable dynamics, we introduce directional dissipative flow restrictors between the input and the inflatable hysterons, as shown in \cref{fig:dynamic}A. We assume that the membranes relax to mechanical equilibria on timescales much faster than the fluidic dynamics, neglecting inertial and viscoelastic effects of membrane material, and all transient behavior arises from fluid transport through the restrictors. To model the transient dynamics, we describe the change in volume of each hysteron as the flow across the restrictor 
\begin{equation}
    \dot{v}_i=
    \begin{cases}
        g_i^+ \, (P_\mathrm{in}-p_i) \quad \mathrm{if} \quad P_\mathrm{in}\geq p_i\\
        g_i^- \, (P_\mathrm{in}-p_i) \quad \mathrm{if} \quad P_\mathrm{in} < p_i,
    \end{cases}
    \label{eq:flow}
\end{equation}
where $g_i^+$ and $g_i^-$ are the forward and backward conductivities of the flow restrictor, respectively. In the quasistatic limit, these dissipative elements do not alter the equilibrium manifold. During dynamic loading, however, they enable temporary pressure differences between the input and the actuators, allowing the system to follow nonequilibrium trajectories. This behavior reflects a separation of memory mechanisms with respect to the input signal: while the discrete hysteron states encode persistent memory through switching events, the dissipative dynamics within a given state exhibit fading memory as the system relaxes toward equilibrium.

% \begin{figure*}
%     \centering
%     \includegraphics[scale=1.0]{FigExp.pdf}
%     \caption{\BG{I don't think the bottom graph of subfigure c adds much. I would integrate the states in the top graph of subfigure c, with zones that show you in which state you are. And then I would add multiple snapshots to show, e.g., some transitions that are not feasible quasi-statically.} Experimental system and validation. \textbf{(A)} Picture of the inflatable hysterons and part of the setup. We apply a constant positive pressure on the upper side of the inflatables to enable the generation of net negative pressures. \textbf{(B)} Measured hysteron pressure hysteretic responses. \textbf{(C)} Example pressure pulses input and measured system state evolution, showing non-trivial transitions. \textbf{(D)} Transition map showing the identified up and down transition zones for each hysteron under a pressure pulse loading. See Supplementary Video for a recording of the experiment.}
%     \label{fig:exp}
% \end{figure*}

The switching time for a hysteron $i$ under a pressure step input $P_\mathrm{in}$ from an initial pressure $p_{i_0}$ can be obtained by integrating \cref{eq:flow}, yielding
% \begin{equation}
%     t_{i_\mathrm{th}}(s_i) = \frac{1}{g_i^\pm \, k_i(s_i)} \,\log\left( \frac{P_\mathrm{in}-p_{i_0}}{P_\mathrm{in}-p_{i_\mathrm{th}}(s_i)} \right).
%     \label{eq:transition}
% \end{equation}
\begin{equation}
\begin{aligned}
t_{i_\mathrm{th}}(s_i)
    &=
    -\tau_i(s_i)\,
    \log\left(
    1-\frac{1}{\eta_i(s_i)}
    \right),
    \quad \eta_i(s_i)>1, \\
\tau_i(s_i)
    &=
    \frac{1}{g_i^\pm \, k_i(s_i)}, \qquad
\eta_i(s_i)
    =
    \frac{P_\mathrm{in}-p_{i_0}}
         {p_{i_\mathrm{th}}(s_i)-p_{i_0}} ,
\end{aligned}
    \label{eq:transition}
\end{equation}
where $t_{i_\mathrm{th}}(s_i)$ is the switching time, $\tau_i(s_i)$ is the characteristic response time of hysteron $i$, and $\eta_i(s_i)$ is a dimensionless driving parameter; $k_i(s_i)$ and $p_{i_\mathrm{th}}(s_i)$ denote the stiffness and switching threshold associated with state $s_i$, while $g_i^\pm$ is the conductivity corresponding to the sign of $(P_\mathrm{in}-p_i)$.
% This switching condition separates into independent pressure and temporal scales. Defining $\tilde{t}= k\, g\, t$ and $\tilde{P}_\mathrm{in} = (P_\mathrm{in}-p_0)/(p_\mathrm{th}-p_0)$), the switching boundary collapses to the universal form $\tilde{t}_\mathrm{th} = \log(\tilde{P}_\mathrm{in}/(\tilde{P}_\mathrm{in}-1))$. Therefore, $k\,g$ and $p_\mathrm{th}-p_0$ independently control the temporal and pressure scales of the control map, providing orthogonal design parameters for shaping transition regions.
Notice that the characteristic response time is set by $(g_i^\pm k_i)^{-1}$, while the characteristic pressure scale is determined by $|p_{i_\mathrm{th}}-p_{i_0}|$, providing orthogonal design variables for tuning the location of the transition regions. Switching only happens if $\eta_i(s_i)>1$, which occurs either when switching up ( $P_\mathrm{in}>p_{i_\mathrm{th}}(s_i)>p_{i_0}$), or when switching down ($P_\mathrm{in}<p_{i_\mathrm{th}}(s_i)<p_{i_0}$).
Under pulse loading, a hysteron switches state only if the pulse amplitude satisfies these conditions and the pulse duration exceeds $t_{i_\mathrm{th}}(s_i)$. 

As an example, we consider the case of three fluidic hysterons connected to the same pressure source, as shown in \cref{fig:dynamic}A, with hysteron characteristics shown in \cref{fig:static}A. In this configuration, there exists a neutral pressure $P_n$ at which all possible states $S$ are stable. Driving the network with pressure pulses around a $P_n$ (initial condition $p_{i}=P_n\,\forall i$) enables transitions that are not accessible under quasistatic loading, as shown in \cref{fig:dynamic}B.
Further, \cref{eq:transition} allows the construction of a two-dimensional control map in the space of pulse amplitude and duration, where regions corresponding to the switching of individual hysterons are identified. For example, the pulse parameters marked by the cross in \cref{fig:dynamic}C, corresponding to the first pulse in \cref{fig:dynamic}B, selectively induce the up-transition of hysteron 2 and thus realize the transition $(0,0,0)\rightarrow(0,1,0)$, which is forbidden under quasistatic loading. 
Since the dynamic evolution of the hysterons is decoupled, these transition boundaries are independent of the discrete system state. For this specific system, these transition regions do not fully overlap, the transition regions do not fully overlap, yielding pulse parameters that selectively address individual hysterons. Therefore, the resulting control map provides independent bit addressability and augments the quasistatic transition graph with dynamically accessible pathways. Sufficient relaxation between successive pulses is required to restore the neutral equilibrium condition and recover the predicted transition boundaries.

\begin{figure}
    \centering
    \includegraphics[scale=1.0]{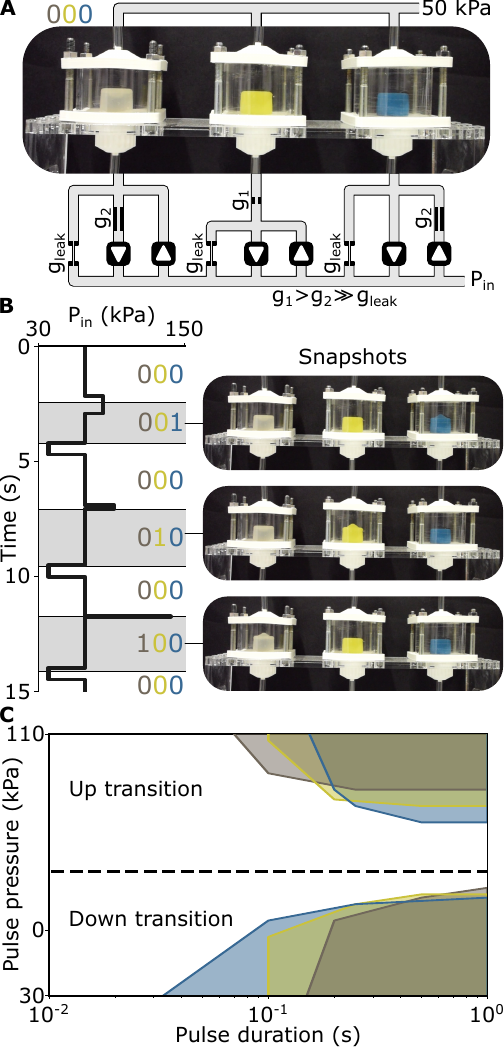}
    \caption{Experimental system and validation. \textbf{(A)} Picture of the inflatable hysterons and schematics of the network. \textbf{(B)} Example pressure pulses input and measured system state evolution, showing non-trivial transitions. \textbf{(C)} Transition map showing the identified up and down transition zones for each hysteron under a pressure pulse loading. See Supplementary Video (\href{https://youtu.be/BmZiTmukhUM}{\texttt{https://youtu.be/BmZiTmukhUM}}) for a recording of the experiment.}
    \label{fig:exp}
\end{figure}

We validate the proposed principle experimentally in a network of three hysteretic disk-spring inflatables~\cite{van2023nonlinear} connected through a dissipative network, as shown in \cref{fig:exp}A. The inflatables share the same geometry but are manufactured utilizing different silicones, resulting in different snapping pressure thresholds. Directional flow restriction is implemented using orifice restrictors and one-way valves, with the effective conductivities tuned by the length of the restrictor in each branch. Additionally, a thin needle restrictor in parallel with the valves is used to guarantee asymptotic equilibrium. The network was loaded with air and driven by a proportional pressure control valve (FESTO, Germany), interfaced with a microcontroller (Arduino, Italy). A constant pressure was applied on the opposite sides of the inflatables, as shown in \cref{fig:exp}A, to enable imposing net negative pressures in the input. Actuating this system with pressure pulses enables individual-bit actuation, as shown in \cref{fig:exp}B, where a train of pulses can selectively switch up each hysteron. The pulse pressure-duration map, in \cref{fig:exp}C, shows individual addressability zones for both up and down transitions, in agreement with the proposed framework. The required pulse profiles, however, differ from the expected values in the quasistatic limit due to the inherent hysteresis in real one‑way valves. 
Moreover, the compressibility of air results in mass accumulation in the tubes, leading to unmodeled dynamics. Accounting for these effects may shift the controllability boundaries, but does not alter the qualitative structure of the pulse design map or the existence of fully disjoint selective transition windows.

This framework of independent single-bit control can be extended to a system with an arbitrary number of hysterons. A general design rule to achieve this necessitates: (i) The forward conductivity of each restrictor should increase with the pressure peak of its associated hysteron. This ensures that sufficiently fast pulses selectively activate elements with higher switching pressures, while elements with lower thresholds do not reach their switching thresholds within the pulse duration. (ii) The backward conductivity should increase as the pressure valley becomes deeper, enabling analogous selectivity during unloading. In particular, when peaks and valleys are nested (the peak pressures follow the reverse ordering of the valleys), symmetric flow restrictors suffice to achieve independent bit control.

In this work, we demonstrate that dynamic loading in dissipative networks enables nontrivial transitions in systems of coupled hysterons, even under a single scalar input. As a result of the time response introduced by the dissipative couplings, the system becomes sensitive not only to the magnitude of the input but also to its temporal evolution. By introducing a simple architecture with effectively decoupled dynamics and a general design rule, we show that transient departures from equilibrium can be harnessed to trigger individual switching events selectively. While static loading can give rise to a large variety of transition graphs~\cite{van2021profusion}, they share a common limitation: transitions are restricted to the direction of the driving field. For a system of $N$ hysterons sharing a common neutral field value ($2^N$ stable states), quasistatic driving yields $2^{(N+1)}-2$ transitions ($2-2^{(1-N)}$ average degree). In contrast, dynamic actuation through a dissipative network enables up to $N\,2^N$ transitions ($N$ average degree), allowing each state to access all single-bit flips and inducing the connectivity of an $N$-dimensional hypercube. This enhanced connectivity guarantees that any state can be reached from any other through at most $N$ transitions, equal to their Hamming distance, eliminating Garden-of-Eden states.

Despite these benefits, the current proposal has some limitations. The effectiveness of our approach relies on a sufficient separation between the hysteron switching thresholds, which determines the separation of the corresponding regions in the pulse amplitude direction. Also, as the number of elements increases, the separation of timescales leads to an increasing difference between the fastest and slowest transitions. Since the experimental setup typically constrains the minimum achievable pulse duration, this sets a practical limit on the overall actuation speed.

Beyond the fully addressable regime considered here, the same dissipative-network design provides flexibility in shaping the accessible transitions. By tuning the conductivities, dynamically enabled transitions can be selectively suppressed while preserving the underlying quasistatic behavior.
More generally, the framework can be extended to coupled systems in which the effective switching thresholds depend on the discrete state. In this regime, the transition of one hysteron can delay that of another, suggesting inhibitory interactions and richer forms of collective dynamics.

Finally, the mechanism described here is not specific to pneumatic systems, but applies more broadly to dissipatively coupled networks of multistable storage elements. In pneumatic networks, multistable inflatables store volume and are driven by pressure, while flow restrictors provide dissipation; this structure is analogous to RC circuits, where charge is stored and driven by voltage through resistive elements. Likewise, multistable mechanical structures store displacement and are driven by force, with dampers providing dissipation (a mechanically equivalent realization is shown in \cref{fig:dynamic}A), analogous to RL circuits, where flux linkage is stored and driven by current through resistive elements. In all such systems, transient departures from equilibrium can create dynamically accessible transition pathways without altering the underlying equilibrium structure.

\begin{acknowledgments}
This work was supported by the European Commission under the Horizon Europe program under Grant \#101076036 (ILUMIS).
\end{acknowledgments}

\onecolumngrid
\clearpage
\begin{center}
  \vspace{12pt}
  \textbf{\large End Mattter}
  \vspace{12pt}
\end{center}
\twocolumngrid

\begin{figure}
    \centering
    \includegraphics[scale=1.0]{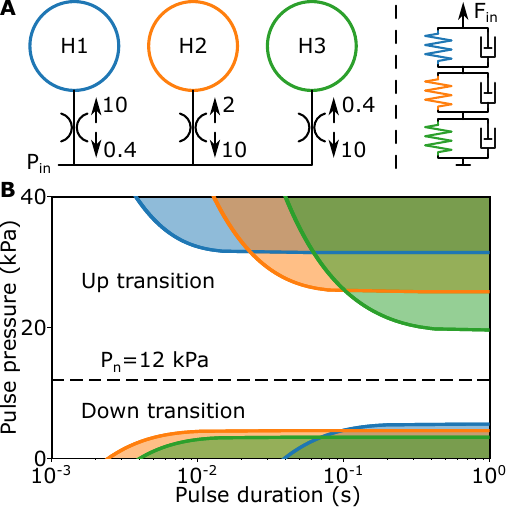}
    \caption{Suppressing dynamic transition in a network. \textbf{(A)} Dissipative network with direction-dependent flow restrictors (conductivity values are in \unit{ml/(s.kPa)}) and the equivalent linear spring system. \textbf{(B)} Control map showing the up and down transition zones for each hysteron under a pressure pulse for a neutral pressure of \SI{12}{kPa}.}
    \label{fig:suppression}
\end{figure}

\textbf{Transition suppression.} 
The proposed dissipative framework also enables the selective suppression of dynamically enabled transitions by tuning the relative timescales of the hysterons through the restrictor conductivities.
For example, increasing the backward conductivity of the connection of hysteron 2 reduces its characteristic deflation time, causing it to respond faster than hysteron 3 during unloading. Consequently, no pulse can snap down hysteron 3 without also snapping down hysteron 2, as reflected in the control map of \cref{fig:suppression}, where the snap-down region of hysteron 3 is entirely contained within that of hysteron 2.
In general, transition suppression is constrained by the ordering of the switching thresholds: the up-transition region of a hysteron with a lower pressure peak cannot be entirely covered by that of a hysteron with a higher pressure peak.

% \begin{figure*}
%     \centering
%     \includegraphics[scale=1.0]{FigCoupled.pdf}
%     \caption{Coupled network of inflatable hysterons. \textbf{(A)} Dissipative network with direction-dependent flow restrictors (conductivity values are in \unit{ml/(s.kPa)}). \textbf{(B, C, D)} State-dependent control maps for saturated \textbf{(B)} and intermediate \textbf{(C)} states showing the up and down transition zones for each hysteron under a pressure pulse for a neutral pressure of \SI{12}{kPa}.}
%     \label{fig:coupled}
% \end{figure*}

\begin{figure}[t]
    \centering
    \includegraphics[scale=1.0]{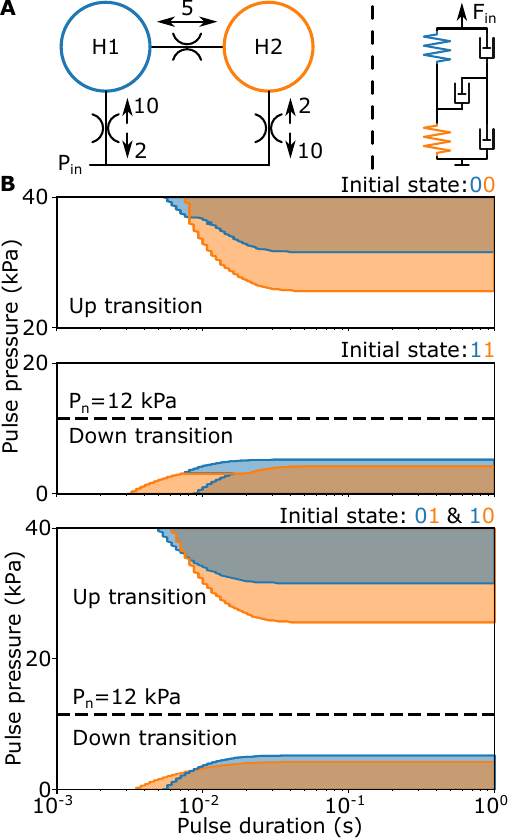}
    \caption{Coupled network of inflatable hysterons. \textbf{(A)} Dissipative network with direction-dependent flow restrictors (conductivity values are in \unit{ml/(s.kPa)}) and the equivalent linear spring system. \textbf{(B)} State-dependent control maps computed for a neutral pressure of \SI{12}{kPa}. The control maps for states $(0,1)$ and $(1,0)$ are superimposed; only one up-transition and one down-transition zone correspond to each state.}
    \label{fig:coupled}
\end{figure}

\textbf{Coupled system.}
We now consider hysterons connected through additional dissipative pathways, as shown in \cref{fig:coupled}A. In this configuration, the dynamics can no longer be described as independent processes, as the evolution of each hysteron depends on the states of the others. These couplings enable effects such as avalanches and transitions during the relaxation phase following a pulse input. Moreover, the outcome of a given pulse depends not only on its parameters but also on the initial discrete state of the system.
This behavior is illustrated in \cref{fig:coupled}B, which reports state-dependent control maps for all four possible initial configurations. Coupling manifests itself through the distortion of these maps; in particular, the transition of one hysteron can delay that of another, suggesting inhibitory interactions. 
These results show that inter-element coupling introduces state-dependent accessibility and richer dynamical behavior than in the decoupled case, but they become more difficult to analyse.

\textbf{Flow control.}
Under imposed flow-rate actuation, the dynamics become coupled even in the absence of direct links between hysterons, since the injected flow must be redistributed among all elements according to their instantaneous states. Consequently, accessibility becomes a coupled dynamical problem, and no simple description in terms of input parameters exists. While any pressure-driven trajectory can in principle be reproduced through an equivalent flow input, the required signals are generally complex, highlighting the practical advantage of pressure control for this class of systems.

\end{document}